\documentclass[aps,preprintnumbers,superscriptaddress,nofootinbib,aps,prd,twocolumn,floatfix]{revtex4-1}
\pdfoutput=1

\usepackage{multirow,array}
\usepackage{amsmath,amssymb}
\usepackage{graphicx}
\usepackage{physics}
\usepackage{color}
\usepackage{url}
\usepackage{slashed}
\usepackage{multirow}
\usepackage{footmisc}
\usepackage{hyperref}
\usepackage{braket}
\usepackage{colortbl}

\hypersetup{
  pdfauthor={Tao Han, Yang Ma, Keping Xie}
  pdftitle={High Energy Leptonic Collisions and Electroweak Parton Distribution Functions}
  pdfkeywords={Electroweak, Parton distribution functions, colliders}
}

\newcommand{\bea}{\begin{eqnarray}}
\newcommand{\eea}{\end{eqnarray}}
\newcommand{\beq}{\begin{equation}}
\newcommand{\eeq}{\end{equation}}

\newcommand{\lsea}{\ell_{\textrm{sea}}}
\newcommand{\ee}{e^+e^-}
\newcommand{\mm}{\mu^+\mu^-}

\begin{document}
\title{High Energy Leptonic Collisions and Electroweak Parton Distribution Functions}

\author{Tao Han}
\email{than@pitt.edu}
\author{Yang Ma} 
\email{mayangluon@pitt.edu}
\author{Keping Xie} 
\email{xiekeping@pitt.edu}
\affiliation{PITT PACC, Department of Physics and Astronomy, University of Pittsburgh, 3941 O'Hara St., Pittsburgh, PA 15260, USA} 

\preprint{PITT-PACC-2006}

%%%%%%%%%%%%%%%%%%%%%%%%%%
\begin{abstract}
 \noindent
In high-energy leptonic collisions well above the electroweak scale, the collinear splitting mechanism of the  electroweak gauge bosons becomes the dominant phenomena via the initial state radiation and the final state showering. We point out that at future high-energy lepton colliders, such as a multi-TeV muon collider, the  electroweak parton distribution functions (EW PDFs) should be adopted as the proper description for partonic collisions of the initial states. 
The leptons and electroweak gauge bosons are the EW partons, that evolve according to the unbroken Standard Model  (SM) gauge group and that effectively resum potentially large collinear logarithms. 
We present a formalism for the EW PDFs at the Leading-Log (LL) accuracy. We calculate semi-inclusive cross sections for some important SM processes at a future multi-TeV muon collider. 
We conclude that it is appropriate to adopt the EW PDF formalism for future high-energy lepton colliders.
\end{abstract}

\maketitle

%%%%%%%%%%%%%%%%%%%%%%%%%%
\noindent
{\bf I. Introduction}\\
%High-energy lepton colliders, such as the Large Electron-Positron Collider (LEP) at CERN and the SLAC Linear Collider (SLC) provided 
With the discovery of the Higgs boson at the CERN Large Hadron Collider (LHC), the particle spectrum of the Standard Model (SM) is complete. The next target at the energy frontier will be to study the Higgs properties  and to search for the next scale beyond the SM \cite{group2019physics}. The physics potential for TeV-scale $\ee$ linear colliders, such as the International Linear Collider \cite{bambade2019international} and the CERN Compact Linear Collider (CLIC) \cite{roloff2018compact}, has been studied to great details. More recently, due to the breakthrough in the cooling technology for a muon beam \cite{delahaye2019muon}, there has been renewed interest in constructing a $\mm$ collider reaching a center-of-momentum energy (c.m.~energy) $\sqrt s \sim {\cal O}(10$ TeV). Advancement of the wake-field electron acceleration technology \cite{collaboration2019advanced} has also been encouraging to have stimulated our ambition for reaching multi-TeV threshold in leptonic collisions. 
%This would be an exciting opportunity to open a new energy threshold of 

Lepton colliders provide a clean experimental environment for precision measurements of physical observables and for discovery of new particles. Near a mass threshold, the $\ee$ annihilation may produce a new particle singly in the $s$-channel, or a particle/anti-particle in pair. As the beam energy increases, the initial state radiation (ISR) becomes substantial. It not only degrades the colliding energies of the leptons, but also generates new reactions of the radiation fields. The most familiar phenomenon is the collinear photon radiation off the high energy charged particles
%, with the leading order approximation 
given by the Weizs\"aicker-Williams spectrum \cite{vonWeizsacker:1934nji,Williams:1934ad} 
%\bea
%\sigma( e^{-}(p) a \to e^{-}(p') X) = {1\over 2S} \overline\Sigma |{\cal M}|^{2} {d\vec p' \over (2\pi)^{3} 2E'}\ dPS_{X} \equiv  \int dx\ P_{\gamma/e}(x)\ \sigma(\gamma a), \\
%\sigma_{EPA}( e^{-} A \to e^{-} X) = \int dx\ {\cal P}_{\gamma/e}(x)\ \hat\sigma(\gamma A),
%\eea
\bea
{\cal P}_{\gamma,\ell}(x)  \approx   \frac{\alpha}{ 2\pi} P_{\gamma,\ell}(x) \ln{E^2\over m^2_\ell}, 
\label{eq:EPA}
\eea
where the splitting functions are $P_{\gamma/\ell}(x) = (1+(1-x)^2)/x$ for $\ell\to \gamma$ 
and $P_{\ell/\ell}(x) = (1+x^2)/(1-x)$ for $\ell\to \ell$, with an energy $xE$ off the charged lepton beam of energy $E$. This is the leading order effective photon approximation (EPA). 

%-------------------------------------------------------
\begin{figure}[tb]
%\vspace{0.6cm}
  \includegraphics[width=.45\textwidth]{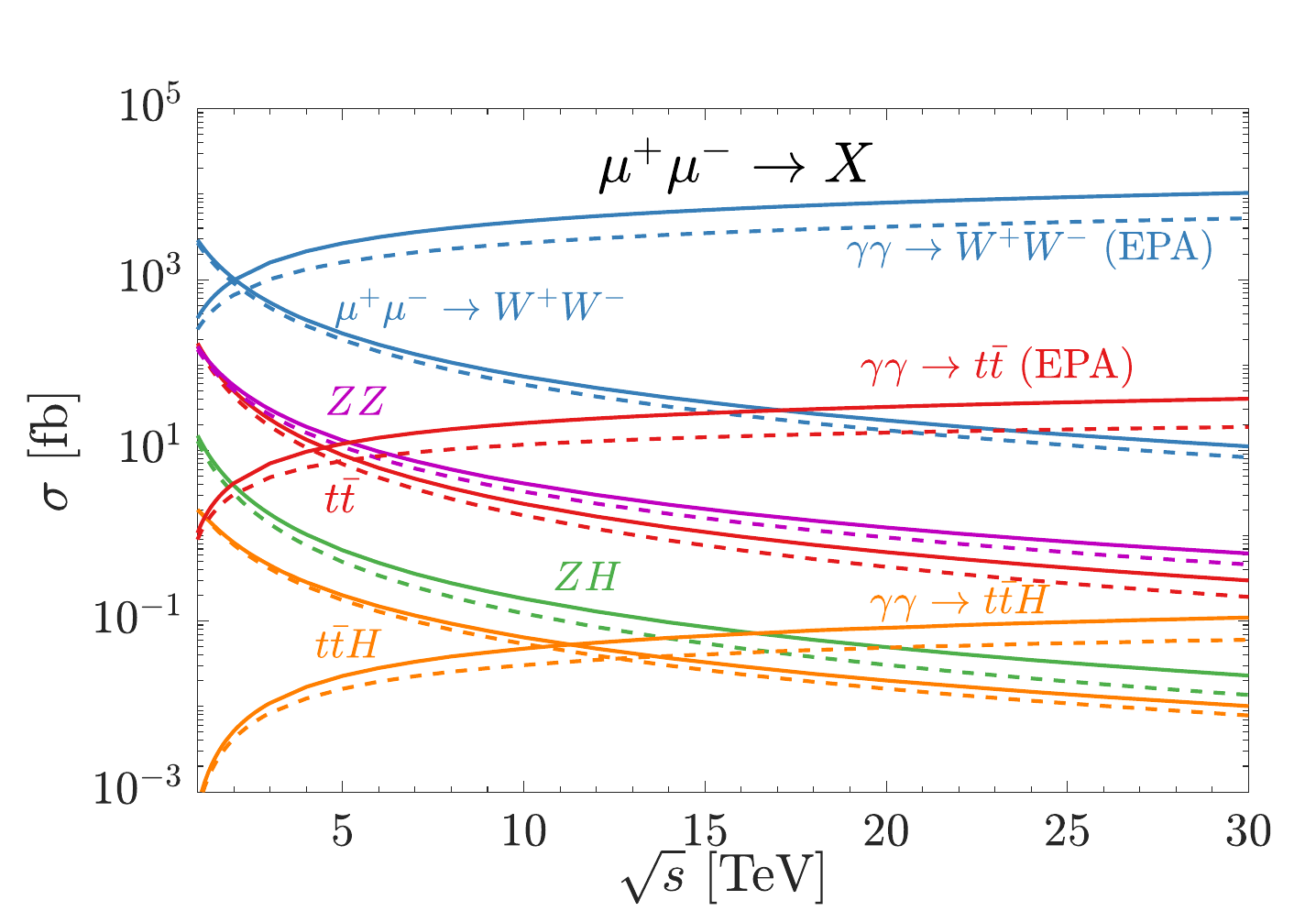}
  %\hspace{0.2cm}
%  \includegraphics[width=.23\textwidth]{diagram3}\\
 %\vspace{0.3cm}
%  \includegraphics[width=.23\textwidth]{diagram5}\hspace{0.2cm}
%  \includegraphics[width=.23\textwidth]{diagram6}
\caption{Production cross sections in $\mm$ collisions versus the c.m.~energy. The dashed falling curves are for the direct $\mm$ annihilation, and the solid falling curves (slightly above the dashed) include the ISR effects. The solid rising curves are for 
$\gamma\gamma$-EPA by Eq.~(\ref{eq:EPA}) and the dashed rising curves are from the leading-order $\gamma$-PDF at $Q=\sqrt{\hat{s}} /2$. 
}
\label{fig:FO}
\end{figure}
%-------------------------------------------------------

In Fig.~\ref{fig:FO}, we show some representative production cross sections 
%at the leading order 
versus the $\mm$ c.m.~energy $\sqrt s$ for
\beq
\mm \to W^+W^-,\ ZZ,\ t \bar t, \ ZH\ {\rm and}\ \ t\bar t H. 
\label{eq:mm}
\eeq
The dashed (falling) curves are for the direct $\mm$ annihilation, and the solid curves (slightly above the dashed) include the ISR effects \cite{Greco_2016}. 
%Here and henceforth, we generically denote those processes as $\mm$ annihilations (ann.). 
We see the typical fall of the annihilation cross sections as $1/s$. 
The ISR reduces the c.m.~energy at the collision and thus increases the cross section. At $\sqrt s = 10$ TeV (30 TeV), the cross section for $\mu^+\mu^- \to t\bar t$ production can be enhanced by 40\% (60\%) due to the ISR effects. 
%The solid rising curves are for $\gamma\gamma$ fusion in the effective photon approximation (EPA) by Eq.~(\ref{eq:EPA}). Those processes take over by orders of magnitude at higher energies due to the collinear logarithmic enhancement. 
%
Owing to the collinear enhancement, the two-photon ($\gamma\gamma$) fusion processes 
grow double-logarithmically. We calculate the total cross sections with the EPA spectrum in Eq.~(\ref{eq:EPA}) for 
\beq
\gamma\gamma \to W^+W^-,\ \ t \bar t \ \ {\rm and}\ \ t\bar t H .
\label{eq:gg}
\eeq 
Those processes take over the annihilation channels at higher energies $\sqrt s \approx 2.5, 4.5, 11$ TeV for $W^+W^-,\ t\bar t$ and $t\bar t H$ production, respectively, 
as shown in Fig.~\ref{fig:FO} by the rising solid curves labeled by EPA. At $\sqrt s \approx 30$ TeV, the production rate for $\gamma\gamma \to t\bar t$ is higher by two orders of magnitude than that for $\mm \to t\bar t$ annihilation. 

However, this description becomes inadequate at some high scales. First, at high energies $E \gg m_\ell$, the collinear logarithm $(\alpha/2\pi) \ln{(E^2 / m^2_\ell)}$ may be sizeable and needs to be resummed for reliable predictions. This leads to the QED analogue of the Dokshitzer-Gribov-Lipatov-Altarelli-Parisi (DGLAP) equations \cite{Dokshitzer:1977sg,Gribov:1972ri,Altarelli:1977zs}, the concept of QED parton distribution functions (PDFs) for the photon and charged fermions \cite{Spiesberger_1995,Roth_2004,Martin_2005}. To estimate the resummation effects, we plot the cross sections with the leading-order $\gamma$-PDF with a scale $Q=\sqrt{\hat s}/2$, where $\sqrt{\hat s}$ is the $\gamma\gamma$ c.m.~energy. As shown in Fig.~\ref{fig:FO} 
by the dashed rising curves below those of EPA, we see that the rates are lowered as expected, and could be smaller by about a factor of two at 30 TeV.
%Furthermore, 

More importantly, as pointed out in Refs.~\cite{Chen:2016wkt,Bauer:2017isx,Bauer:2018arx} and explored in details \cite{ours}, at scales $Q^2 \gg M_Z^2$, the SM gauge symmetry SU(2)$_L \otimes$U(1)$_Y$ is effectively restored. Consequently, the four EW gauge bosons ($W^{\pm,3},B$) in the SM must be taken into account all together coherently with $B$-$W^3$ mixing and interference. The fermion interactions are chiral and the couplings and states evolve according to the SM unbroken gauge symmetry. One needs to invoke the picture of electroweak parton distribution functions (EW PDFs) \cite{Ciafaloni_2000,Ciafaloni_2002,Manohar:2018kfx} dynamically generated by the electroweak and Yukawa interactions. The longitudinally polarized gauge bosons capture the remnants of the EW symmetry breaking. The effects are governed by power corrections of the order $M_Z^2/Q^2$ \cite{Kane:1984bb,Dawson:1984gx}, a measure of the Goldstone-Boson Equivalence violation \cite{Chen:2016wkt,Cuomo:2019siu}, analogous to higher-twist effects in QCD. 

\begin{figure}[bt]
%\vspace{0.6cm}
 \includegraphics[width=.42\textwidth]{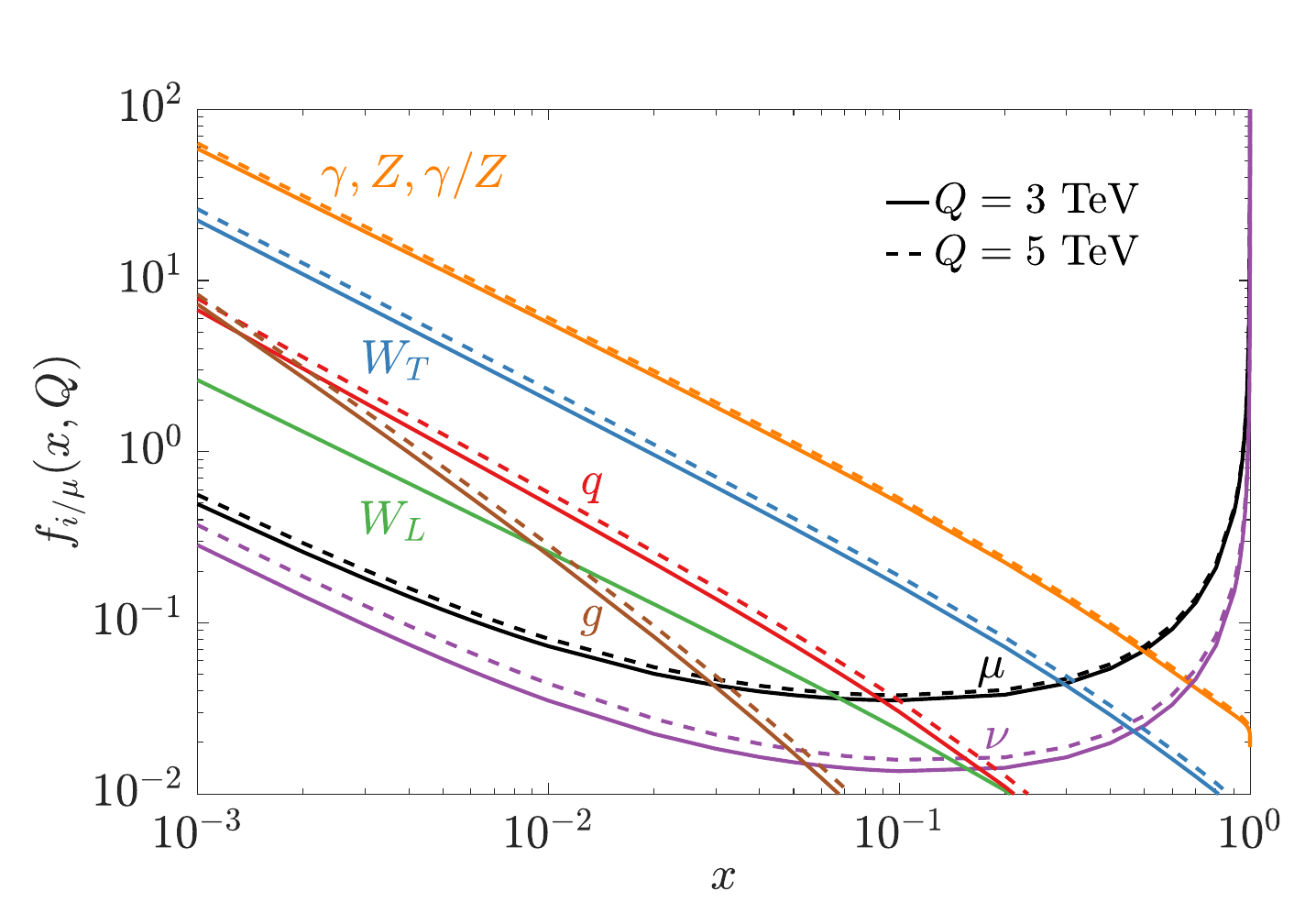}
  \includegraphics[width=.42\textwidth]{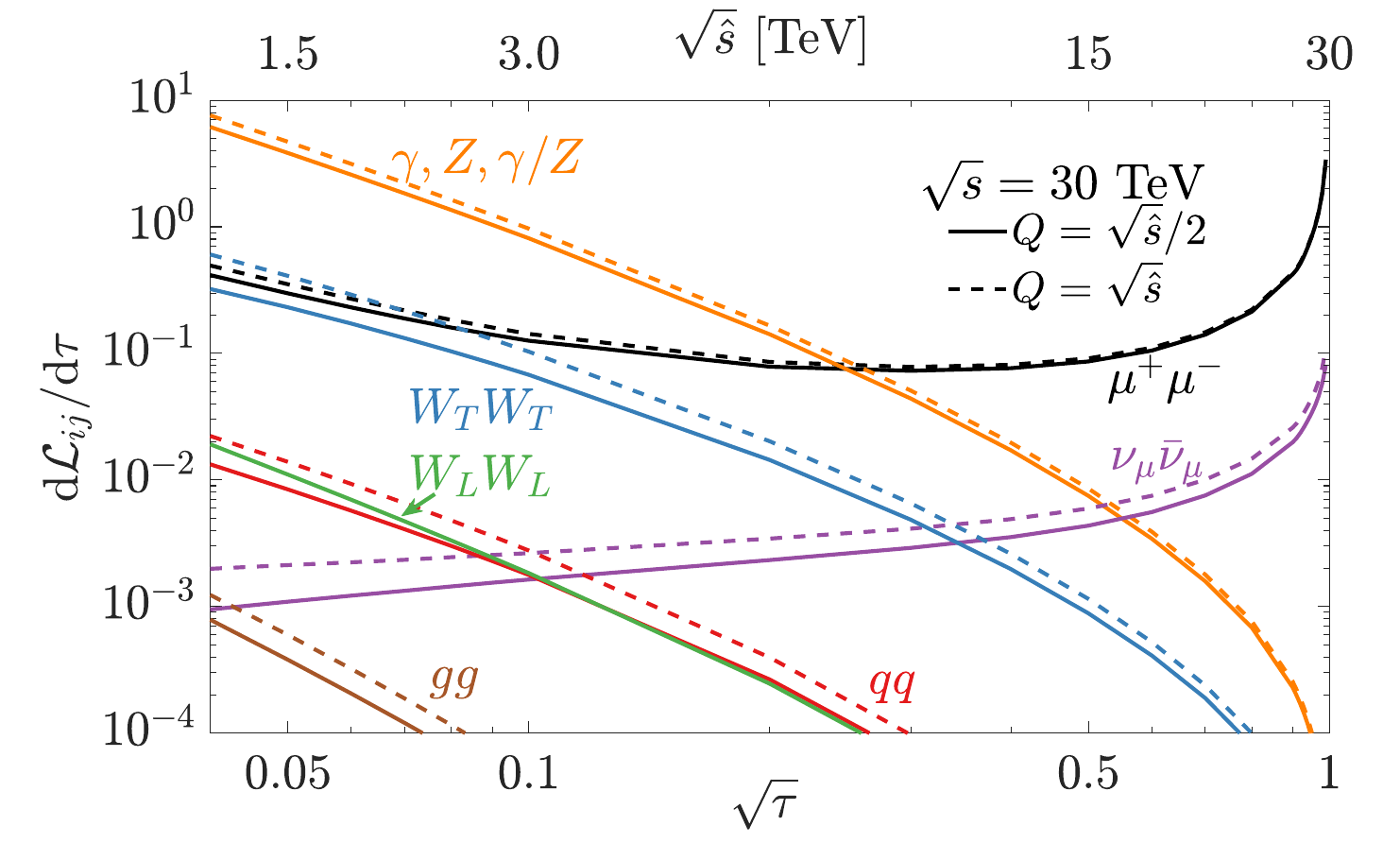}
\caption{Distributions for (a) EW PDFs $f_i(x)$ and, (b) parton luminosities $\dd\mathcal{L}_{ij}/\dd\tau$ versus 
$\sqrt \tau$ for $\sqrt s = 30$ TeV with a factorization scale $Q=\sqrt{\hat s}/2$ (solid) and $\sqrt{\hat s}$ (dashed). }
\label{fig:L}
\end{figure}

%%%%%%%%%%%%%%%%%%%%%%%%%%%%%%%%%%%%%
\vskip 0.1cm
\noindent
{\bf II. Electroweak Parton Distribution Functions}\\  
Below the EW scale $Q< M_Z$, the effects of the SU(2)$_L$ gauge bosons are suppressed by $g^2/M_Z^2$. The gauge boson radiation off a charged lepton beam ($\ell^\pm=e^\pm, \mu^\pm$) is essentially purely electromagnetic.
%, the photon ($\gamma$). 
%Thus the lepton and the photon ($\gamma$) can be viewed as ``electromagnetic partons'' \cite{QED}. 
At the EW scale and above, all electroweak states in the unbroken SM  are dynamically activated. The massless states involved at the leading order are
\beq
\ell_R,\ \ell_L,\ \nu_L \ \ {\rm and} \ \ B, W^{\pm,3}.
\label{eq:states}
\eeq 
We will not include the Higgs sector in the initial state partons since the Yukawa couplings to $e,\mu$ are not relevant for the current consideration.
% $(y_\mu \sim 6\times 10^{-4})$. 
However, we must include the effects of longitudinally polarized gauge bosons 
%are the messengers to retain the remnant effect of the electroweak symmetry breaking, 
characterized by power corrections of the order $M_Z^2/Q^2$. 
Denote an EW PDF as $f_i(x,Q^2)$ with $i$ labeling a particle with an energy fraction $x$ at a factorization scale $Q$. 
The EW PDFs evolve according to the full EW DGLAP equations \cite{Ciafaloni:2005fm,Bauer:2017isx}
\beq
{\dd f_{i} \over \dd\ln Q^2}=\sum_{I}\frac{\alpha_I}{2\pi}\sum_{j}P^{I}_{i,j}\otimes f_{j},
\eeq
where $I$ specifies the gauge group, and the $P_{ij}^{I}$ are the splitting functions for $j\to i$. The complete list of the EW splitting functions for the SM chiral states are available in Refs.~\cite{Ciafaloni_2002,Chen:2016wkt,Bauer:2017isx}. 
The initial condition for a lepton beam is $f_\ell(x, m_\ell^2)\approx \delta(1-x) + {\cal O}(\alpha)$ and it evolves as $\ln(Q^2/m_\ell^2)$. At the electroweak scale, the matching conditions are $f_\gamma(x,M_Z^2)\ne 0,\ \ f_Z(x,M_Z^2) = 0,\ \ f_{\gamma Z}(x,M_Z^2) = 0$, with a general relation 
\begin{equation}\label{eq:rotation}
\begin{pmatrix}
f_B\\ f_{W^3}\\ f_{BW^3}
\end{pmatrix}=\begin{pmatrix}
c_W^2 & s_W^2  & -c_W s_W\\
s_W^2 & c_W^2  & c_W s_W\\
2c_W s_W & -2c_W s_W & c_W^2-s_W^2 
\end{pmatrix}
\begin{pmatrix}
f_\gamma \\ f_Z\\ f_{\gamma Z}
\end{pmatrix},
\nonumber
\end{equation}
where $s_W=\sin\theta_W$ is the weak mixing angle. The mixed PDF $f_{\gamma Z}$ (or $f_{BW^3}$) represents a mix state and is important to account for the interference between the diagrams involving $\gamma/Z$ (or $B/W^3$) \cite{Ciafaloni_2000,Chen:2016wkt,Bauer:2017isx}. Chiral couplings and their RGE running are fully taken into account including the correlation between the polarized PDFs and the corresponding polarized scattering amplitudes. With one-loop virtual corrections, our results are accurate at the leading-log (LL) order. In Fig.~\ref{fig:L}(a), we present EW PDFs for the states in Eq.~(\ref{eq:states}) for $\ell=\mu$ with a scale $Q=$3 TeV and 5 TeV. For completeness, we have also included the quarks $q=\sum_{i=d}^t(q_i+\bar{q}_i)$ and gluons from the higher-order splittings.
We give the averaged momentum fractions $\langle xf_i\rangle=\int xf_i(x)\dd x$ carried by various parton species in Table \ref{tab:mom}. 
The two scale choices lead to less than $20\%$ difference for the EW PDFs. As expected, the fermionic states sharply peak at $x\approx 1$, while the bosonic states peak at $x\approx 0$, reflecting the infrared behavior. It is noted that there is an enhanced rate at small $x$ for the fermions, deviating from the leading order behavior $\sim 1/(1-x)$. This is from the soft $\gamma^*/Z^*/W^*$ splitting at higher orders. Owing to the large flux of photons at low scales, the neutral EW PDFs are largest. Unlike all the other EW PDFs that scale logarithmically with $Q$, the longitudinal gauge bosons $(W_L,Z_L)$ do not scale with $Q$ at the leading order \cite{Ciafaloni_2001,Chen:2016wkt,Bauer:2017isx} $-$ an explicit example for Bjorken-scaling restoration. 
%We have not included the Higgs boson and the Goldstone bosons. Although their PDFs all run logarithmically, the Yukawa couplings to the leptons are highly suppressed.
% If we are interested in some subtle physics effects of the order $(M_Z^2/\hat{s})\sim 10^{-4}$, the muon Yukawa contribution may become relevant. 
\begin{table}
  \begin{tabular}{c|c|c|c|c|c|c|c}
  \hline
  $Q$ & $\mu$ &$\gamma,Z,\gamma Z$ & $W^\pm$ & $\nu$  & $\lsea$ & $q$ & $g$  \\
  \hline
  $M_Z$ & 97.9 & 2.06 & 0  & 0  & 0.028  & 0.035  & 0.0062  \\
  3 TeV  & 91.5 &  3.61 & 1.10   & 3.59  & 0.069  & 0.13  & 0.019   \\
  5 TeV  & 89.9  &  3.82 & 1.24  & 4.82 & 0.077   & 0.16  & 0.022 \\
  \hline
  \end{tabular}
  \caption{Momentum fractions (\%) carried by various parton species. The sea leptons include $\lsea=\bar{\mu}+\sum_{i\neq\mu}(\ell_i+\bar{\ell}_i)$ and $\nu=\sum_{i}(\nu_i+\bar{\nu}_i)$. The quark components include all the 6 flavors.}
  %: $q=\sum_{i}(q_i+\bar{q}_i)$.}
  \label{tab:mom}
\end{table}

%-------------------------------------------------------
  
%%%%%%%%%%%%%%%%%%%%%%%%%%%%%%%%%%%%
\vskip 0.1cm
\noindent
{\bf III. Cross sections for Semi-inclusive Processes in $\mm$ Collisions}\\  
We write the production cross section of an exclusive final state $F$ and the unspecified remnants $X$  
in terms of the parton luminosity $\dd\mathcal{L}_{ij}/{\dd\tau}$ and the corresponding partonic sub-process cross section $ \hat{\sigma}$
\begin{eqnarray}
&  \sigma(\ell^+\ell^- \rightarrow F + X) 
%&=&  \sum_{i,j}
%  \int^{1}_{\tau_{0}}d\xi_{a}\int^{1}_{\tau_{0}\over \xi_a}d\xi_{b} 
%  \left[f_{i/p}(\xi_a,Q_{f}^{2})f_{j/p}(\xi_b,Q_{f}^{2}) \hat{\sigma}(i j \rightarrow X) + (i\leftrightarrow j)\right]~~ \label{factTheorem.EQ}  \\   &=& 
= \int_{\tau_{0}}^{1} \dd\tau  \sum_{ij}\frac{\dd\mathcal{L}_{ij}}{\dd\tau}\  \hat{\sigma}(ij\rightarrow F),
\label{eq:fact}  \\ 
& {\dd\mathcal{L}_{ij} \over {\dd\tau} } = \frac{1}{1+\delta_{ij}}  \int^{1}_{\tau} \frac{\dd\xi}{\xi} 
 \left[ f_{i}(\xi, Q^{2})f_{j}\left(\frac{\tau}{\xi},Q^{2} \right) + (i \leftrightarrow j) \right], 
 \nonumber
\end{eqnarray}
where 
%$\xi_{a,b}$ are the fractions of momenta carried by initial partons $(i,j)$, 
%$Q$ is the factorization scale, 
$\tau = \hat s/s$ with $\sqrt{s}\ (\sqrt{\hat{s}})$ the collider (parton) c.m.~energy. The production threshold is $\tau_{0} = m_{F}^{2}/s$.

In presenting our results for production of SM particles at a high-energy lepton collider, for definitiveness, we consider a future $\mm$ collider with multi-TeV energies. It is informative to first examine the parton luminosities as shown in Fig.~\ref{fig:L}(b) for $\sqrt s=30$ TeV versus $\sqrt \tau$, with a variety of partonic initial states. The upper horizontal axis labels the accessible $\sqrt{\hat s}$. Although we properly evolve the EW PDFs according to the unbroken SM gauge groups, we convert the states back for the sake of common intuition, shown in the figure for $\mm,\ \nu_\mu \bar \nu_\mu,\ \gamma\gamma/ZZ/\gamma Z,\ W_TW_T$ and $ W_LW_L$. We see that the fermionic luminosities peak near the machine c.m.~energy $\tau\approx 1$, while the gauge boson luminosities, generically called vector boson fusion (VBF) dominate at lower partonic energy $\sqrt{\hat s}$. As noted earlier, the neutral gauge boson luminosities are the largest, followed by $W_T$ and $W_L$. 
% For instance, assuming a 14 TeV muon collider, the partonic luminosities for these two initial state combinations cross near 3 TeV, as marked on the upper scale of Fig.~\ref{fig:L}(b). 

We emphasize the ``inclusiveness'' of the production processes. For example, for an exclusive final state of $t \bar t$ production, one needs to sum over all the observationally indistinguishable partonic contributions in the initial state $\mm, \gamma\gamma, \gamma Z, ZZ, W^+W^- \to t\bar t$. 
Contributions from the quark and gluon initial states are sub-leading 
%suppressed by more than 3 orders of magnitude  due to the small 
as seen in the parton luminosities in Fig.~\ref{fig:L}(b), and 
% with respect to the $\gamma\gamma$ initiated sub-processes. For this reason, 
we do not include them in the cross section calculations throughout this letter. 
Since the collinear remnants are not observationally resolved, one cannot separate the $\mm/ \nu_\mu \bar \nu_\mu$ annihilations from the VBF. For this reason, we call such processes, {\it i.e.}, $\mm \to t\bar t$ ``semi-inclusive''. This is analogous to the $t\bar t$ production at hadron colliders from the partonic sub-processes $q\bar q, gg \to t\bar t$. 

\begin{figure}[t]
%\vspace{0.6cm}
%  \includegraphics[width=.48\textwidth]{figure/ttbar_inclusive_log.pdf}
%   \includegraphics[width=.238\textwidth]{figure/sigma_PDF.pdf}
%   \includegraphics[width=.238\textwidth]{figure/sigma_tt.pdf}
   \includegraphics[width=.45\textwidth]{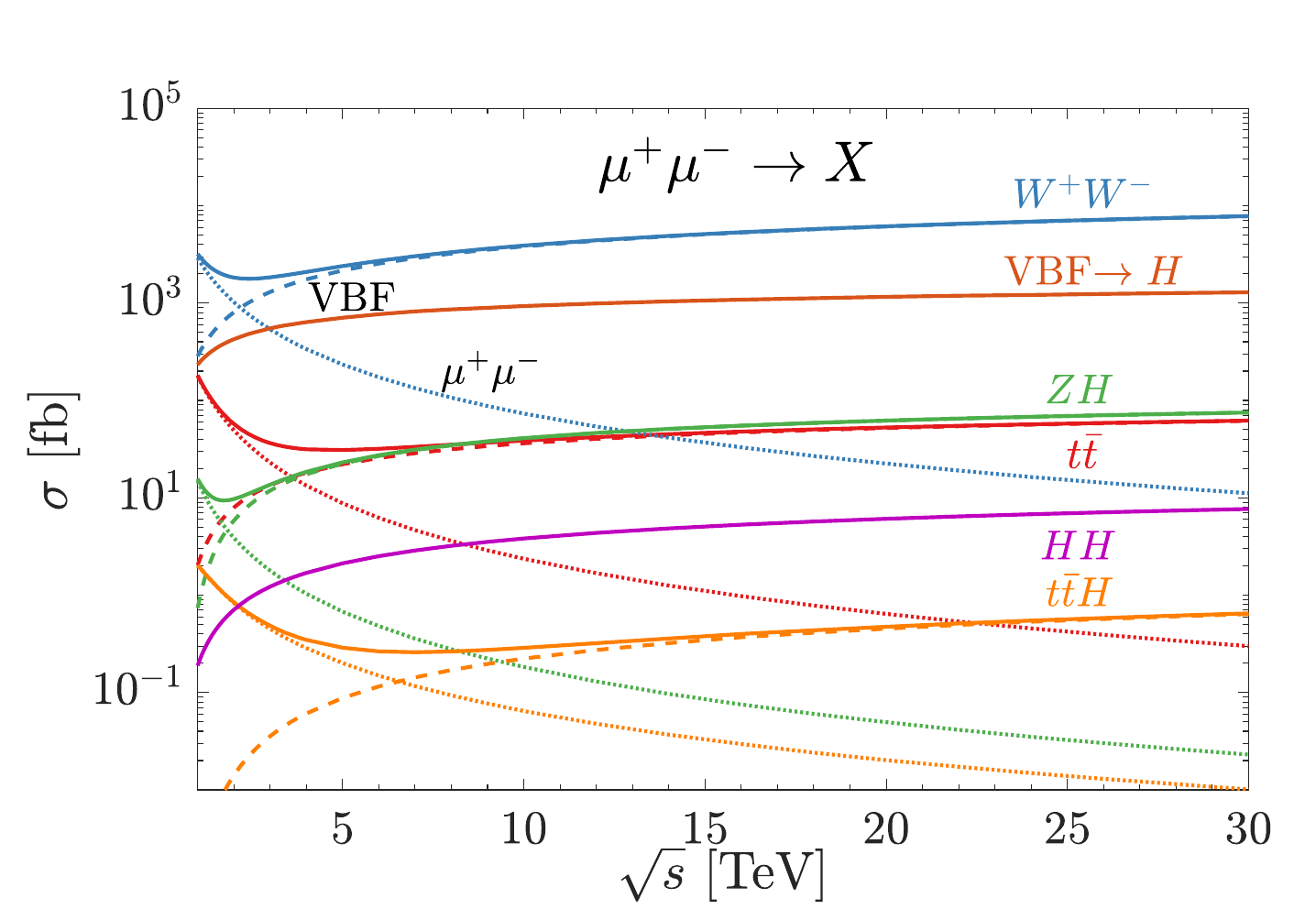}
   \includegraphics[width=.45\textwidth]{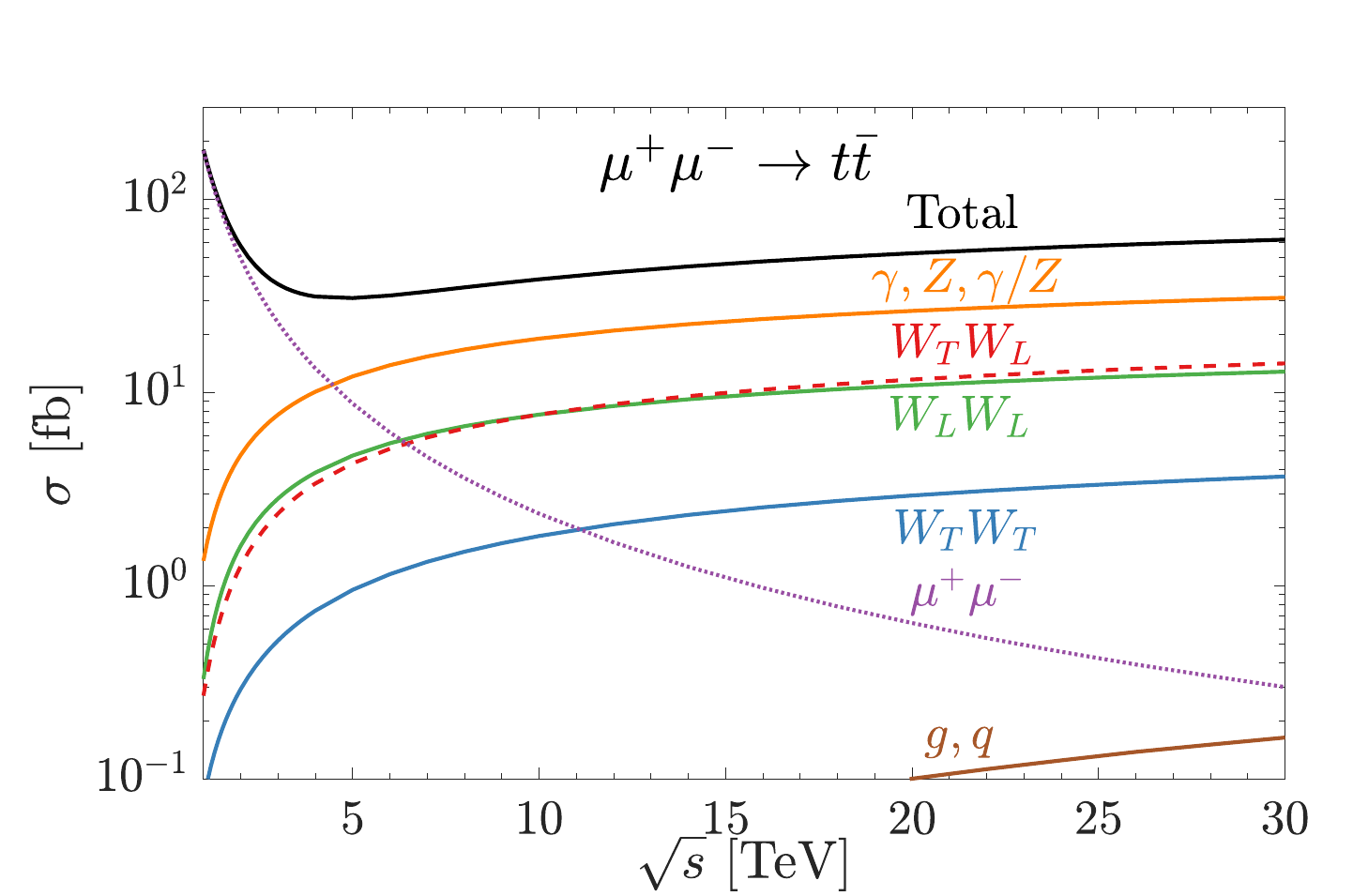}
  %\hspace{0.2cm}
%  \includegraphics[width=.23\textwidth]{diagram6}
\caption{Production cross section for semi-inclusive processes at a $\mm$ collider versus the c.m.~energy. The solid curves are for (a) the total cross sections and the dashed (dotted) curves from VBF ($\mm$ annihilation) with EW PDF, and (b) for $t \bar t$ production decomposed to the underlying contributions from $\mm, \gamma/Z/\gamma Z, W_TW_L$, $W_LW_L$ and $W_TW_T$. 
%Dotted curves indicate the  calculations from $\gamma\gamma$-PDF and FO $WW$ fusion. 
}
\label{fig:semi}
\end{figure}

In Fig.~\ref{fig:semi}(a), we show the semi-inclusive production cross sections at a $\mm$ collider versus the collider c.m.~energy $\sqrt{s}$ from 1 TeV to 30 TeV. We choose the factorization scale $Q=\sqrt{\hat s}/2$ in calculating the EW PDFs.\footnote{To validate the EW PDF approximation,
% and to regularize the collinear behavior for the partonic scattering processes, 
we have imposed an angular cutoff for the $W/Z$ initiated processes in the c.m.~frame $\cos\theta < 1-m^2/\hat{s}$, where $m$ is the relevant particle mass involved in the process. 
%The $W^+W^-, ZH$ final state is singular due to the $t$-channel photon exchange. 
We have included a tighter cut $\cos\theta < 0.99$ and $\sqrt{\hat s} > 500$ GeV for the $W^+W^-, ZH$ final states.}
The solid curves are the total cross sections for the semi-inclusive processes for
%as in Eq.~(\ref{eq:mm}), including the Higgs production $H$ and $HH$. 
%
\beq
\mm \to W^+W^-,\ H,\ ZH,\ t \bar t, \ HH\ \ {\rm and}\ \ t\bar t H ,
\label{eq:mm}
\eeq
combining the contributions from both fermionic initial states and the VBF. We indicate the VBF contributions by the dashed curves,\footnote{Many of the VBF processes have been calculated recently in Ref.~\cite{costantini2020vector}
at the tree-level. We have good agreements with theirs where ever they overlap.} and the fermionic contributions by the dotted curves, respectively, below the solid curves. It is important to note that, although there is no logarithmic evolution for the $W_L$ PDF, the partonic sub-process cross sections are much enhanced for $W_LW_L,Z_LZ_L \to t\bar t, t\bar t H$ and $H, ZH, HH$, due to the Goldstone-boson interactions. 
The VBF processes take over the annihilation channels at higher energies $\sqrt s \approx 2.3, 3.5, 6.5$ TeV for $W^+W^-,\ t\bar t$ and $t\bar t H$, respectively. 
To appreciate the individual contributions from the underlying partonic subprocesses, we decompose them for the process $\mm \to t\bar t$ versus the c.m.~energy, as shown in Fig.~\ref{fig:semi}(b) for 
$\mm$, $\gamma\gamma/\gamma Z/ZZ$, $W_TW_L,\ W_LW_L$ as well as $W_TW_T$. As expected, the QED contribution remains to be the leading channel. Not well appreciated, the $W_TW_L/W_LW_L$ contributions become as significant. 
%
%We first see that the gross features agree between the EPA and the EW PDF. However, there is a significant enhancement for the EW PDF treatment. 
%
%For comparison, we also depict by the dotted curves from the results of $\gamma$-PDF and the FO $WW$ fusion \cite{costantini2020vector}. We see that our predictions for the cross sections with the EW PDFs at $\sqrt s = 10$ TeV (30 TeV) can be different from those by a factor $15\%\ (20\%)$. for ... 
% For certain processes, this could lead to a reduction in rate of $20\% - 50\%$. }

%-------------------------------------------------------
%-------------------------------------------------------

\begin{figure}[t]
%\vspace{0.6cm}
%  \includegraphics[width=.238\textwidth]{figure/tt_dmtt.pdf}
%    \includegraphics[width=.23\textwidth]{figure/tt_dytt.pdf}
  \includegraphics[width=.42\textwidth]{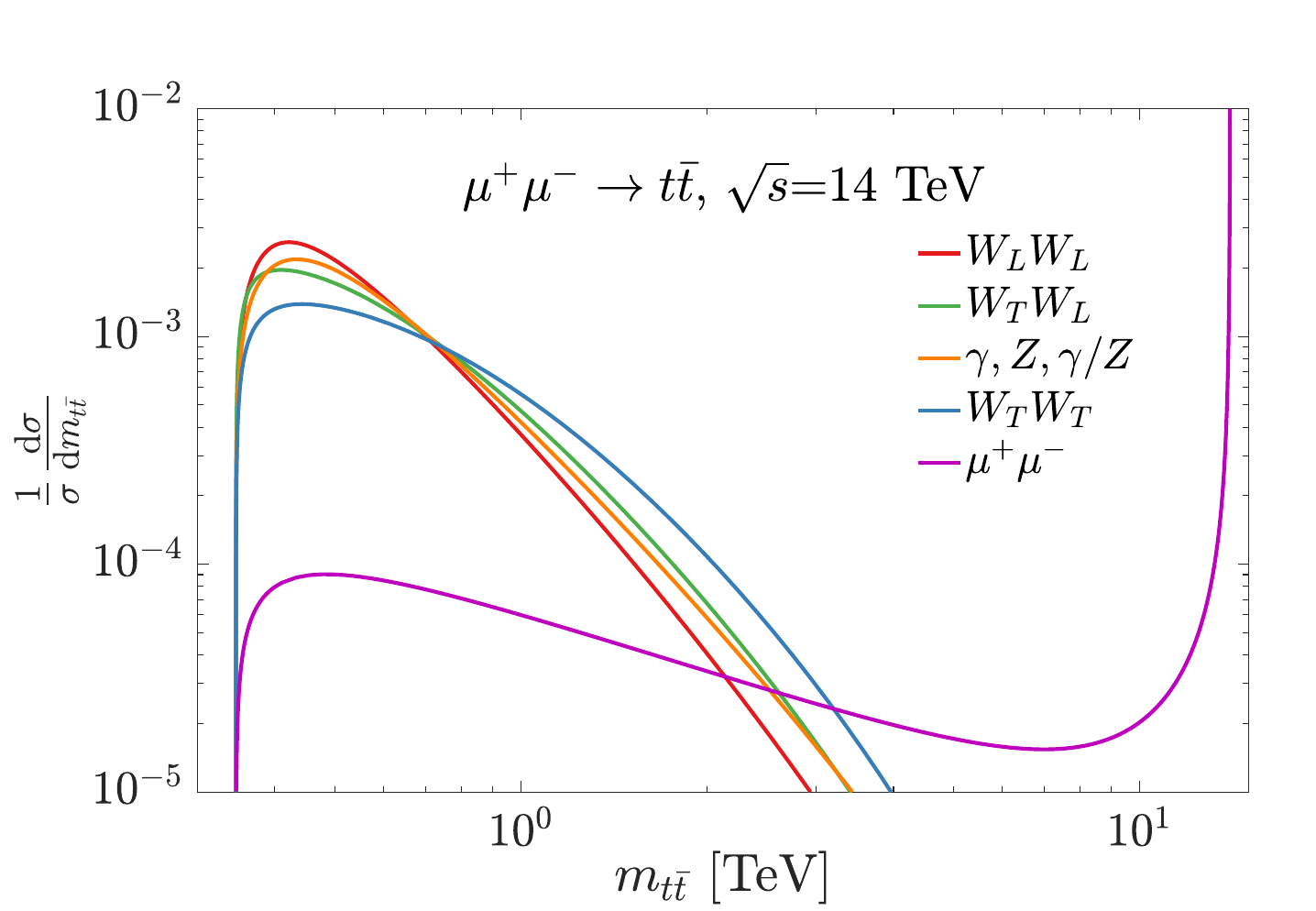}
    \includegraphics[width=.42\textwidth]{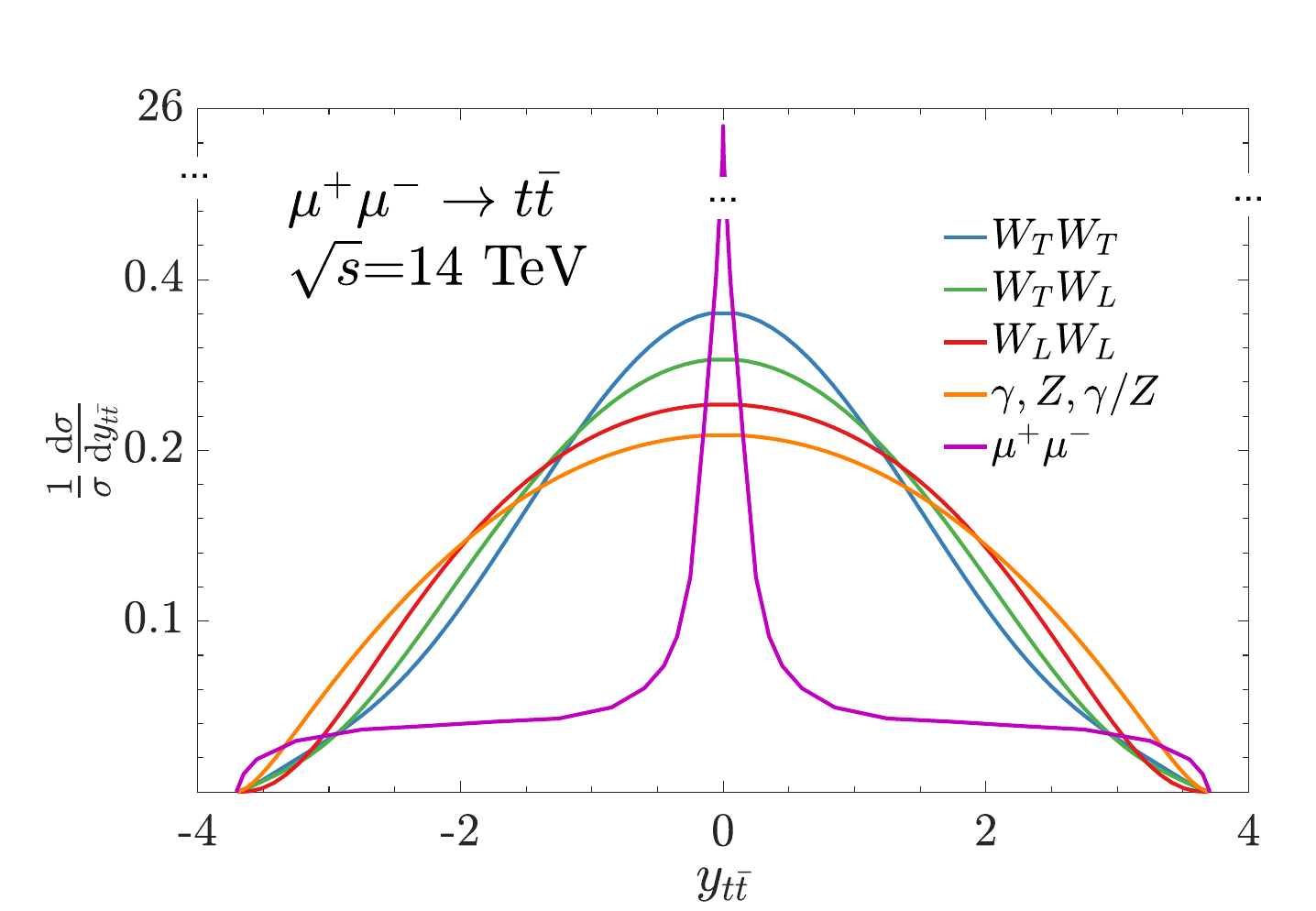}
  %\hspace{0.2cm}
%  \includegraphics[width=.23\textwidth]{diagram6}
\caption{Normalized differential distributions for the final state $t\bar t$ system (a) the invariant mass $m_{t \bar t}$ and (b) the rapidity $y_{t\bar t}$. }
\label{fig:inv}
\end{figure}

We now examine the kinematic distributions for the final state $t\bar t$ system, for the individual contributions $\mm, \gamma/Z, W_TW_L, W_LW_L$ and $W_TW_T$. Shown in Fig.~\ref{fig:inv}(a) are the normalized invariant mass distributions $m_{t \bar t}$. We see that, for the $\mm$ annihilation, the distribution is sharply peaked at the collider c.m.~energy, with a tail due to the radiative return. For the VBF, they are peaked after the $2 m_t$ threshold.  
We show in Fig.~\ref{fig:inv}(b) the normalized rapidity distributions of the system $y_{t \bar t}$. Again, events from the $\mm$ annihilation are sharply central, while those from VBF are spread out, reflecting 
%These distributions reflect the underlying partonic reaction: the invariant mass is essentially the $\sqrt{\tau}$ spectrum convoluted with the dynamical matrix elements, while the rapidity is 
the boost due to the momentum imbalance between the two incoming partons.

%-------------------------------------------------------
%-------------------------------------------------------

We summarize our results utilizing the EW PDFs in Table \ref{tab:xsec} for a few characteristic processes for a muon collider with a few representative energies 3, 6, 10, 14 and 30 TeV. For the sake of illustration, we once again separate the partonic sub-processes by the fermionic annihilation and by VBF.

\begin{table*}[htb]
\centering
\begin{tabular}{|c|cc|cc|cc|cc|cc|}
\hline
   $\sqrt s$ (TeV)       & \multicolumn{2}{c|}{3}  & \multicolumn{2}{c|}{6} & \multicolumn{2}{c|}{10} & \multicolumn{2}{c|}{14} & \multicolumn{2}{c|}{30} \\
   %\vspace{1mm} \\
    $\sigma$ (fb)& VBF &$\mu\mu$  & VBF &$\mu\mu$ & VBF &$\mu\mu$  & VBF&$\mu\mu$ & VBF&$\mu\mu$  \\
\hline
$W^+W^-$ &    1300 & 540    & 2500  & 170  &  3800 & 73              &  4900 & 41 &   7800 & 11  \\
%$ZZ$  &   1/6     &      &   1/2              &  1 &   2/3  \\
    $t\bar t$  &   13 &  23    & 25  &  6.2    &  36 &   2.4            & 43 & 1.3  &  61 & 0.30   \\
    \hline
    $ZH$  &   12 & 1.8   &  26 &  0.48 &     41 &  0.18             & 51 & 0.097 &  75 & 0.023  \\   
%    $h$  &  1      &   0   &           &   0                &    \\
   $HH$ &   1.2 &--     & 2.5  &  --    & 3.8 &-- &  4.8 &-- &  7.6&--         \\         
        $t\bar t H$  &   0.036 & 0.45    &  0.12 &  0.15       & 0.22& 0.065    &   0.32  & 0.037 & 0.64& 0.010 \\
       \hline
\end{tabular}
\caption{Production cross sections at a muon collider in units of fb by VBF utilizing the EW PDF and by direct $\mm$ annihilation with ISR effects. }
\label{tab:xsec}
\end{table*}

%%%%%%%%%%%%%%%%%%%%%%%%%%%%%%%%%%%%%
\vskip 0.1cm
\noindent
{\bf IV. Discussions and Conclusion}
%Several remarks are in order.

\noindent 
$\bullet$ 
The naive EPA is inadequate at high scales. The QED evolution of $\ln(Q^2/m_\ell^2)$ in the $\gamma$-PDF should capture the dominant effect at an appropriate physical scale $Q^2$. Although the $Z$ contribution is typically small until reaching a very high scale, the mixed state $\gamma Z (BW^3)$ needs to be taken into account that often interferes destructively. 

\noindent
$\bullet$ 
%Last but not least, 
The EW PDF approach allows for calculating individual contributions from the polarized initial state partons, with correlations to the corresponding sub-process matrix elements. This is an important feature when polarization is needed for exploring a certain type of underlying dynamics. This option would be unavailable with the fixed order (FO) diagrammatic calculations \cite{Alwall:2014hca,Kilian:2007gr,costantini2020vector}. In addition, the FO calculations 
%is not only a conceptual advancement for the treatment of high energy leptonic collisions, but also 
%provides a substantial simplification for VBF processes over the fixed order diagrammatic calculations \cite{Alwall:2014hca,Kilian:2007gr} where one 
may face a tremendous challenge for numerical stability dealing with the large collinear logs $\ln(Q^2/m_\ell^2)$. 
%But be aware of $\cos\theta < 1-m^2/\hat{s}$.

\noindent
$\bullet$ 
%In the unbroken phase of the SM, Goldstone bosons decouple from the light fermions to the accuracy of muon Yukawa coupling $\sim 6\times 10^{-4}$. 
%The longitudinally polarized gauge bosons are the messengers to keep the remnant effect of the electroweak symmetry breaking, characterized by power corrections of the order $M_Z^2/Q^2$. 
Although no logarithmic growth for the longitudinally polarized gauge boson PDFs, the large Yukawa coupling to the top quark and the scalar self-interaction of the Goldstone bosons make the sub-processes substantially enhanced, as seen for the VBF production of $t\bar t, t\bar t H, ZH$ and $HH$.

%\noindent
%$\bullet$ \kp{When solving the DGLAP equations iteratively, we could reach a percentage accuracy by five interactions for weak-gauge-boson ($\gamma/Z/W$) PDFs. Physically, this corresponds to a showering process with consecutive collinear radiations. The convergence for the lepton, quark, and gluon PDFs is poor due to the $\log^p(x)$ terms in the $x\to1$ limit. We need to reach $\mathcal{O}(40)$ iterative steps to obtain a good convergence.} 

\noindent 
$\bullet$ 
For the PDFs of fermions with a bare SU(2) charge, due to the incomplete cancellation of the infrared divergence, they are not exactly factorizable. 
%\footnote{Caution must be taken here that, unlike QED and QCD, the factorization of Eq.~(\ref{eq:fact}) for a fermion with a bare SU(2) charge does not hold due to the incomplete cancellation of the soft singularity.
%double-log dependence on the high scale. 
This is known as the violation of the Bloch-Nordsieck theorem \cite{Ciafaloni_2002,Chen:2016wkt}. 
%We adopt a simple cutoff $\tau^{max}=1-M_Z^2/\hat{s}$ to regularize the double-log behavior. 
% \cite{Bauer:2017isx}. 
%This introduces mild logarithmic sensitivity to the cutoff \cite{ours}. }
%
This does not pose a problem to the beam (valence) lepton because it is a numerically small 
higher-order correction. This could lead to an unphysical solution to the dynamically generated neutrinos. We impose an infrared cutoff as a regulator $\tau^{\max}=1-M_Z/Q$, which assures the resummation to a double-log accuracy \cite{Bauer:2017isx}. 

\noindent
$\bullet$ 
We have not taken into account the effects of the final-state radiations (FSR). This could become one of the dominant features at very high energies, properly described by the ``fragmentation functions'' \cite{Chen:2016wkt,Bauer:2018xag}. We leave this topic for future explorations.

\noindent
$\bullet$ 
We did not quantify the potentially large corrections near the threshold $Q^2 \ge 4 m^2$. On the one hand, our formalism aims to address the physics far above the threshold $Q^2 \gg M_Z^2$. On the other hand, the infrared behavior of the gauge boson radiation tends to populate the events in the low-$Q^2$ region. We leave this topic for future investigations. 

\noindent
$\bullet$ 
Although we focused on a $\mm$ collider in our presentation, the EW PDF formalism is equally applicable for $\ee$ colliders. The only difference is the QED radiation effects, further enhanced by a factor $\ln(m_\mu^2/m_e^2)$. It is also straightforward to apply our formalism to the high energy hadron colliders, although the photon PDF of the proton at a low scale is more subtle \cite{Manohar:2016nzj,Fornal:2018znf}. 

In summary, we advocated that all particles accessible under the SM interactions should be viewed as EW partons in high energy leptonic collisions well above the EW scale. We presented a systematic approach to define the EW PDFs for leptons and gauge bosons accurate to the order of LL under the unbroken gauge symmetry. We calculated the production cross sections for some characteristic SM processes at a high-energy muon collider in the EW PDF formalism. Polarized partonic cross sections can be evaluated individually that are desirable for exploring new physics beyond the Standard Model at future high energy colliders. 
%The PDFs for the beam muons and photon radiation can lead to effects of about a factor of two. 

%%%%%%%%%%%%%%%%%%%%%%%%%%
\vskip 0.1cm
\textbf{Acknowledgment:}
We thank Christian Bauer, John Collins, Fabio Maltoni, Dave Soper and Bryan Webber for helpful discussions. 
This work was supported in part by the U.S.~Department of Energy under grant No.~DE-FG02- 95ER40896, U.S.~National Science Foundation under Grant No.~PHY-1820760, and in part by the PITT PACC. 
  
%%%%%%%%%%%%%%%%%%%%%%%%%%

\bibliographystyle{apsrev}
\bibliography{ref.bib}

\begin{thebibliography}{32}
\expandafter\ifx\csname natexlab\endcsname\relax\def\natexlab#1{#1}\fi
\expandafter\ifx\csname bibnamefont\endcsname\relax
  \def\bibnamefont#1{#1}\fi
\expandafter\ifx\csname bibfnamefont\endcsname\relax
  \def\bibfnamefont#1{#1}\fi
\expandafter\ifx\csname citenamefont\endcsname\relax
  \def\citenamefont#1{#1}\fi
\expandafter\ifx\csname url\endcsname\relax
  \def\url#1{\texttt{#1}}\fi
\expandafter\ifx\csname urlprefix\endcsname\relax\def\urlprefix{URL }\fi
\providecommand{\bibinfo}[2]{#2}
\providecommand{\eprint}[2][]{\url{#2}}

\bibitem[{\citenamefont{for Particle Physics
  Preparatory~Group}(2019)}]{group2019physics}
\bibinfo{author}{\bibfnamefont{E.~S.} \bibnamefont{for Particle Physics
  Preparatory~Group}}, \emph{\bibinfo{title}{Physics briefing book}}
  (\bibinfo{year}{2019}), \eprint{1910.11775}.

\bibitem[{\citenamefont{Bambade et~al.}(2019)\citenamefont{Bambade, Barklow,
  Behnke, Berggren, Brau, Burrows, Denisov, Faus-Golfe, Foster, Fujii
  et~al.}}]{bambade2019international}
\bibinfo{author}{\bibfnamefont{P.}~\bibnamefont{Bambade}},
  \bibinfo{author}{\bibfnamefont{T.}~\bibnamefont{Barklow}},
  \bibinfo{author}{\bibfnamefont{T.}~\bibnamefont{Behnke}},
  \bibinfo{author}{\bibfnamefont{M.}~\bibnamefont{Berggren}},
  \bibinfo{author}{\bibfnamefont{J.}~\bibnamefont{Brau}},
  \bibinfo{author}{\bibfnamefont{P.}~\bibnamefont{Burrows}},
  \bibinfo{author}{\bibfnamefont{D.}~\bibnamefont{Denisov}},
  \bibinfo{author}{\bibfnamefont{A.}~\bibnamefont{Faus-Golfe}},
  \bibinfo{author}{\bibfnamefont{B.}~\bibnamefont{Foster}},
  \bibinfo{author}{\bibfnamefont{K.}~\bibnamefont{Fujii}},
  \bibnamefont{et~al.}, \emph{\bibinfo{title}{The international linear
  collider: A global project}} (\bibinfo{year}{2019}), \eprint{1903.01629}.

\bibitem[{\citenamefont{Roloff et~al.}(2018)\citenamefont{Roloff, Franceschini,
  Schnoor, and Wulzer}}]{roloff2018compact}
\bibinfo{author}{\bibfnamefont{P.}~\bibnamefont{Roloff}},
  \bibinfo{author}{\bibfnamefont{R.}~\bibnamefont{Franceschini}},
  \bibinfo{author}{\bibfnamefont{U.}~\bibnamefont{Schnoor}}, \bibnamefont{and}
  \bibinfo{author}{\bibfnamefont{A.}~\bibnamefont{Wulzer}},
  \emph{\bibinfo{title}{The compact linear e$^+$e$^-$ collider (clic): Physics
  potential}} (\bibinfo{year}{2018}), \eprint{1812.07986}.

\bibitem[{\citenamefont{Delahaye et~al.}(2019)\citenamefont{Delahaye, Diemoz,
  Long, Mansouli, Pastrone, Rivkin, Schulte, Skrinsky, and
  Wulzer}}]{delahaye2019muon}
\bibinfo{author}{\bibfnamefont{J.~P.} \bibnamefont{Delahaye}},
  \bibinfo{author}{\bibfnamefont{M.}~\bibnamefont{Diemoz}},
  \bibinfo{author}{\bibfnamefont{K.}~\bibnamefont{Long}},
  \bibinfo{author}{\bibfnamefont{B.}~\bibnamefont{Mansouli}},
  \bibinfo{author}{\bibfnamefont{N.}~\bibnamefont{Pastrone}},
  \bibinfo{author}{\bibfnamefont{L.}~\bibnamefont{Rivkin}},
  \bibinfo{author}{\bibfnamefont{D.}~\bibnamefont{Schulte}},
  \bibinfo{author}{\bibfnamefont{A.}~\bibnamefont{Skrinsky}}, \bibnamefont{and}
  \bibinfo{author}{\bibfnamefont{A.}~\bibnamefont{Wulzer}},
  \emph{\bibinfo{title}{Muon colliders}} (\bibinfo{year}{2019}),
  \eprint{1901.06150}.

\bibitem[{\citenamefont{collaboration}(2019)}]{collaboration2019advanced}
\bibinfo{author}{\bibfnamefont{A.}~\bibnamefont{collaboration}},
  \emph{\bibinfo{title}{Towards an advanced linear international collider}}
  (\bibinfo{year}{2019}), \eprint{1901.10370}.

\bibitem[{\citenamefont{von Weizsacker}(1934)}]{vonWeizsacker:1934nji}
\bibinfo{author}{\bibfnamefont{C.~F.} \bibnamefont{von Weizsacker}},
  \bibinfo{journal}{Z. Phys.} \textbf{\bibinfo{volume}{88}},
  \bibinfo{pages}{612} (\bibinfo{year}{1934}).

\bibitem[{\citenamefont{Williams}(1934)}]{Williams:1934ad}
\bibinfo{author}{\bibfnamefont{E.~J.} \bibnamefont{Williams}},
  \bibinfo{journal}{Phys. Rev.} \textbf{\bibinfo{volume}{45}},
  \bibinfo{pages}{729} (\bibinfo{year}{1934}).

\bibitem[{\citenamefont{Greco et~al.}(2016)\citenamefont{Greco, Han, and
  Liu}}]{Greco_2016}
\bibinfo{author}{\bibfnamefont{M.}~\bibnamefont{Greco}},
  \bibinfo{author}{\bibfnamefont{T.}~\bibnamefont{Han}}, \bibnamefont{and}
  \bibinfo{author}{\bibfnamefont{Z.}~\bibnamefont{Liu}},
  \bibinfo{journal}{Physics Letters B} \textbf{\bibinfo{volume}{763}},
  \bibinfo{pages}{409} (\bibinfo{year}{2016}), ISSN \bibinfo{issn}{0370-2693},
  \urlprefix\url{http://dx.doi.org/10.1016/j.physletb.2016.10.078}.

\bibitem[{\citenamefont{Dokshitzer}(1977)}]{Dokshitzer:1977sg}
\bibinfo{author}{\bibfnamefont{Y.~L.} \bibnamefont{Dokshitzer}},
  \bibinfo{journal}{Sov. Phys. JETP} \textbf{\bibinfo{volume}{46}},
  \bibinfo{pages}{641} (\bibinfo{year}{1977}).

\bibitem[{\citenamefont{Gribov and Lipatov}(1972)}]{Gribov:1972ri}
\bibinfo{author}{\bibfnamefont{V.~N.} \bibnamefont{Gribov}} \bibnamefont{and}
  \bibinfo{author}{\bibfnamefont{L.~N.} \bibnamefont{Lipatov}},
  \bibinfo{journal}{Sov. J. Nucl. Phys.} \textbf{\bibinfo{volume}{15}},
  \bibinfo{pages}{438} (\bibinfo{year}{1972}), \bibinfo{note}{[Yad.
  Fiz.15,781(1972)]}.

\bibitem[{\citenamefont{Altarelli and Parisi}(1977)}]{Altarelli:1977zs}
\bibinfo{author}{\bibfnamefont{G.}~\bibnamefont{Altarelli}} \bibnamefont{and}
  \bibinfo{author}{\bibfnamefont{G.}~\bibnamefont{Parisi}},
  \bibinfo{journal}{Nucl. Phys.} \textbf{\bibinfo{volume}{B126}},
  \bibinfo{pages}{298} (\bibinfo{year}{1977}).

\bibitem[{\citenamefont{Spiesberger}(1995)}]{Spiesberger_1995}
\bibinfo{author}{\bibfnamefont{H.}~\bibnamefont{Spiesberger}},
  \bibinfo{journal}{Physical Review D} \textbf{\bibinfo{volume}{52}},
  \bibinfo{pages}{4936} (\bibinfo{year}{1995}), ISSN \bibinfo{issn}{0556-2821},
  \urlprefix\url{http://dx.doi.org/10.1103/PhysRevD.52.4936}.

\bibitem[{\citenamefont{Roth and Weinzierl}(2004)}]{Roth_2004}
\bibinfo{author}{\bibfnamefont{M.}~\bibnamefont{Roth}} \bibnamefont{and}
  \bibinfo{author}{\bibfnamefont{S.}~\bibnamefont{Weinzierl}},
  \bibinfo{journal}{Physics Letters B} \textbf{\bibinfo{volume}{590}},
  \bibinfo{pages}{190} (\bibinfo{year}{2004}), ISSN \bibinfo{issn}{0370-2693},
  \urlprefix\url{http://dx.doi.org/10.1016/j.physletb.2004.04.009}.

\bibitem[{\citenamefont{Martin et~al.}(2005)\citenamefont{Martin, Roberts,
  Stirling, and Thorne}}]{Martin_2005}
\bibinfo{author}{\bibfnamefont{A.~D.} \bibnamefont{Martin}},
  \bibinfo{author}{\bibfnamefont{R.~G.} \bibnamefont{Roberts}},
  \bibinfo{author}{\bibfnamefont{W.~J.} \bibnamefont{Stirling}},
  \bibnamefont{and} \bibinfo{author}{\bibfnamefont{R.~S.}
  \bibnamefont{Thorne}}, \bibinfo{journal}{The European Physical Journal C}
  \textbf{\bibinfo{volume}{39}}, \bibinfo{pages}{155} (\bibinfo{year}{2005}),
  ISSN \bibinfo{issn}{1434-6052},
  \urlprefix\url{http://dx.doi.org/10.1140/epjc/s2004-02088-7}.

\bibitem[{\citenamefont{Chen et~al.}(2017)\citenamefont{Chen, Han, and
  Tweedie}}]{Chen:2016wkt}
\bibinfo{author}{\bibfnamefont{J.}~\bibnamefont{Chen}},
  \bibinfo{author}{\bibfnamefont{T.}~\bibnamefont{Han}}, \bibnamefont{and}
  \bibinfo{author}{\bibfnamefont{B.}~\bibnamefont{Tweedie}},
  \bibinfo{journal}{JHEP} \textbf{\bibinfo{volume}{11}}, \bibinfo{pages}{093}
  (\bibinfo{year}{2017}), \eprint{1611.00788}.

\bibitem[{\citenamefont{Bauer et~al.}(2017)\citenamefont{Bauer, Ferland, and
  Webber}}]{Bauer:2017isx}
\bibinfo{author}{\bibfnamefont{C.~W.} \bibnamefont{Bauer}},
  \bibinfo{author}{\bibfnamefont{N.}~\bibnamefont{Ferland}}, \bibnamefont{and}
  \bibinfo{author}{\bibfnamefont{B.~R.} \bibnamefont{Webber}},
  \bibinfo{journal}{JHEP} \textbf{\bibinfo{volume}{08}}, \bibinfo{pages}{036}
  (\bibinfo{year}{2017}), \eprint{1703.08562}.

\bibitem[{\citenamefont{Bauer and Webber}(2019)}]{Bauer:2018arx}
\bibinfo{author}{\bibfnamefont{C.~W.} \bibnamefont{Bauer}} \bibnamefont{and}
  \bibinfo{author}{\bibfnamefont{B.~R.} \bibnamefont{Webber}},
  \bibinfo{journal}{JHEP} \textbf{\bibinfo{volume}{03}}, \bibinfo{pages}{013}
  (\bibinfo{year}{2019}), \eprint{1808.08831}.

\bibitem[{\citenamefont{Han et~al.}(2020)\citenamefont{Han, Ma, and
  Xie}}]{ours}
\bibinfo{author}{\bibfnamefont{T.}~\bibnamefont{Han}},
  \bibinfo{author}{\bibfnamefont{Y.}~\bibnamefont{Ma}}, \bibnamefont{and}
  \bibinfo{author}{\bibfnamefont{K.}~\bibnamefont{Xie}} (\bibinfo{year}{2020}),
  \eprint{Work in progress}.

\bibitem[{\citenamefont{Ciafaloni et~al.}(2000)\citenamefont{Ciafaloni,
  Ciafaloni, and Comelli}}]{Ciafaloni_2000}
\bibinfo{author}{\bibfnamefont{M.}~\bibnamefont{Ciafaloni}},
  \bibinfo{author}{\bibfnamefont{P.}~\bibnamefont{Ciafaloni}},
  \bibnamefont{and} \bibinfo{author}{\bibfnamefont{D.}~\bibnamefont{Comelli}},
  \bibinfo{journal}{Physical Review Letters} \textbf{\bibinfo{volume}{84}},
  \bibinfo{pages}{4810} (\bibinfo{year}{2000}), ISSN \bibinfo{issn}{1079-7114},
  \urlprefix\url{http://dx.doi.org/10.1103/PhysRevLett.84.4810}.

\bibitem[{\citenamefont{Ciafaloni et~al.}(2002)\citenamefont{Ciafaloni,
  Ciafaloni, and Comelli}}]{Ciafaloni_2002}
\bibinfo{author}{\bibfnamefont{M.}~\bibnamefont{Ciafaloni}},
  \bibinfo{author}{\bibfnamefont{P.}~\bibnamefont{Ciafaloni}},
  \bibnamefont{and} \bibinfo{author}{\bibfnamefont{D.}~\bibnamefont{Comelli}},
  \bibinfo{journal}{Physical Review Letters} \textbf{\bibinfo{volume}{88}}
  (\bibinfo{year}{2002}), ISSN \bibinfo{issn}{1079-7114},
  \urlprefix\url{http://dx.doi.org/10.1103/PhysRevLett.88.102001}.

\bibitem[{\citenamefont{Manohar and Waalewijn}(2018)}]{Manohar:2018kfx}
\bibinfo{author}{\bibfnamefont{A.~V.} \bibnamefont{Manohar}} \bibnamefont{and}
  \bibinfo{author}{\bibfnamefont{W.~J.} \bibnamefont{Waalewijn}},
  \bibinfo{journal}{JHEP} \textbf{\bibinfo{volume}{08}}, \bibinfo{pages}{137}
  (\bibinfo{year}{2018}), \eprint{1802.08687}.

\bibitem[{\citenamefont{Kane et~al.}(1984)\citenamefont{Kane, Repko, and
  Rolnick}}]{Kane:1984bb}
\bibinfo{author}{\bibfnamefont{G.~L.} \bibnamefont{Kane}},
  \bibinfo{author}{\bibfnamefont{W.~W.} \bibnamefont{Repko}}, \bibnamefont{and}
  \bibinfo{author}{\bibfnamefont{W.~B.} \bibnamefont{Rolnick}},
  \bibinfo{journal}{Phys. Lett.} \textbf{\bibinfo{volume}{148B}},
  \bibinfo{pages}{367} (\bibinfo{year}{1984}).

\bibitem[{\citenamefont{Dawson}(1985)}]{Dawson:1984gx}
\bibinfo{author}{\bibfnamefont{S.}~\bibnamefont{Dawson}},
  \bibinfo{journal}{Nucl. Phys.} \textbf{\bibinfo{volume}{B249}},
  \bibinfo{pages}{42} (\bibinfo{year}{1985}).

\bibitem[{\citenamefont{Cuomo et~al.}(2020)\citenamefont{Cuomo, Vecchi, and
  Wulzer}}]{Cuomo:2019siu}
\bibinfo{author}{\bibfnamefont{G.}~\bibnamefont{Cuomo}},
  \bibinfo{author}{\bibfnamefont{L.}~\bibnamefont{Vecchi}}, \bibnamefont{and}
  \bibinfo{author}{\bibfnamefont{A.}~\bibnamefont{Wulzer}},
  \bibinfo{journal}{SciPost Phys.} \textbf{\bibinfo{volume}{8}},
  \bibinfo{pages}{078} (\bibinfo{year}{2020}), \eprint{1911.12366}.

\bibitem[{\citenamefont{Ciafaloni and Comelli}(2005)}]{Ciafaloni:2005fm}
\bibinfo{author}{\bibfnamefont{P.}~\bibnamefont{Ciafaloni}} \bibnamefont{and}
  \bibinfo{author}{\bibfnamefont{D.}~\bibnamefont{Comelli}},
  \bibinfo{journal}{JHEP} \textbf{\bibinfo{volume}{11}}, \bibinfo{pages}{022}
  (\bibinfo{year}{2005}), \eprint{hep-ph/0505047}.

\bibitem[{\citenamefont{Ciafaloni et~al.}(2001)\citenamefont{Ciafaloni,
  Ciafaloni, and Comelli}}]{Ciafaloni_2001}
\bibinfo{author}{\bibfnamefont{M.}~\bibnamefont{Ciafaloni}},
  \bibinfo{author}{\bibfnamefont{P.}~\bibnamefont{Ciafaloni}},
  \bibnamefont{and} \bibinfo{author}{\bibfnamefont{D.}~\bibnamefont{Comelli}},
  \bibinfo{journal}{Physical Review Letters} \textbf{\bibinfo{volume}{87}}
  (\bibinfo{year}{2001}), ISSN \bibinfo{issn}{1079-7114},
  \urlprefix\url{http://dx.doi.org/10.1103/PhysRevLett.87.211802}.

\bibitem[{\citenamefont{Costantini et~al.}(2020)\citenamefont{Costantini,
  Lillo, Maltoni, Mantani, Mattelaer, Ruiz, and Zhao}}]{costantini2020vector}
\bibinfo{author}{\bibfnamefont{A.}~\bibnamefont{Costantini}},
  \bibinfo{author}{\bibfnamefont{F.~D.} \bibnamefont{Lillo}},
  \bibinfo{author}{\bibfnamefont{F.}~\bibnamefont{Maltoni}},
  \bibinfo{author}{\bibfnamefont{L.}~\bibnamefont{Mantani}},
  \bibinfo{author}{\bibfnamefont{O.}~\bibnamefont{Mattelaer}},
  \bibinfo{author}{\bibfnamefont{R.}~\bibnamefont{Ruiz}}, \bibnamefont{and}
  \bibinfo{author}{\bibfnamefont{X.}~\bibnamefont{Zhao}},
  \emph{\bibinfo{title}{Vector boson fusion at multi-tev muon colliders}}
  (\bibinfo{year}{2020}), \eprint{2005.10289}.

\bibitem[{\citenamefont{Alwall et~al.}(2014)\citenamefont{Alwall, Frederix,
  Frixione, Hirschi, Maltoni, Mattelaer, Shao, Stelzer, Torrielli, and
  Zaro}}]{Alwall:2014hca}
\bibinfo{author}{\bibfnamefont{J.}~\bibnamefont{Alwall}},
  \bibinfo{author}{\bibfnamefont{R.}~\bibnamefont{Frederix}},
  \bibinfo{author}{\bibfnamefont{S.}~\bibnamefont{Frixione}},
  \bibinfo{author}{\bibfnamefont{V.}~\bibnamefont{Hirschi}},
  \bibinfo{author}{\bibfnamefont{F.}~\bibnamefont{Maltoni}},
  \bibinfo{author}{\bibfnamefont{O.}~\bibnamefont{Mattelaer}},
  \bibinfo{author}{\bibfnamefont{H.~S.} \bibnamefont{Shao}},
  \bibinfo{author}{\bibfnamefont{T.}~\bibnamefont{Stelzer}},
  \bibinfo{author}{\bibfnamefont{P.}~\bibnamefont{Torrielli}},
  \bibnamefont{and} \bibinfo{author}{\bibfnamefont{M.}~\bibnamefont{Zaro}},
  \bibinfo{journal}{JHEP} \textbf{\bibinfo{volume}{07}}, \bibinfo{pages}{079}
  (\bibinfo{year}{2014}), \eprint{1405.0301}.

\bibitem[{\citenamefont{Kilian et~al.}(2011)\citenamefont{Kilian, Ohl, and
  Reuter}}]{Kilian:2007gr}
\bibinfo{author}{\bibfnamefont{W.}~\bibnamefont{Kilian}},
  \bibinfo{author}{\bibfnamefont{T.}~\bibnamefont{Ohl}}, \bibnamefont{and}
  \bibinfo{author}{\bibfnamefont{J.}~\bibnamefont{Reuter}},
  \bibinfo{journal}{Eur. Phys. J.} \textbf{\bibinfo{volume}{C71}},
  \bibinfo{pages}{1742} (\bibinfo{year}{2011}), \eprint{0708.4233}.

\bibitem[{\citenamefont{Bauer et~al.}(2018)\citenamefont{Bauer, Provasoli, and
  Webber}}]{Bauer:2018xag}
\bibinfo{author}{\bibfnamefont{C.~W.} \bibnamefont{Bauer}},
  \bibinfo{author}{\bibfnamefont{D.}~\bibnamefont{Provasoli}},
  \bibnamefont{and} \bibinfo{author}{\bibfnamefont{B.~R.}
  \bibnamefont{Webber}}, \bibinfo{journal}{JHEP} \textbf{\bibinfo{volume}{11}},
  \bibinfo{pages}{030} (\bibinfo{year}{2018}), \eprint{1806.10157}.

\bibitem[{\citenamefont{Manohar et~al.}(2016)\citenamefont{Manohar, Nason,
  Salam, and Zanderighi}}]{Manohar:2016nzj}
\bibinfo{author}{\bibfnamefont{A.}~\bibnamefont{Manohar}},
  \bibinfo{author}{\bibfnamefont{P.}~\bibnamefont{Nason}},
  \bibinfo{author}{\bibfnamefont{G.~P.} \bibnamefont{Salam}}, \bibnamefont{and}
  \bibinfo{author}{\bibfnamefont{G.}~\bibnamefont{Zanderighi}},
  \bibinfo{journal}{Phys. Rev. Lett.} \textbf{\bibinfo{volume}{117}},
  \bibinfo{pages}{242002} (\bibinfo{year}{2016}), \eprint{1607.04266}.

\bibitem[{\citenamefont{Fornal et~al.}(2018)\citenamefont{Fornal, Manohar, and
  Waalewijn}}]{Fornal:2018znf}
\bibinfo{author}{\bibfnamefont{B.}~\bibnamefont{Fornal}},
  \bibinfo{author}{\bibfnamefont{A.~V.} \bibnamefont{Manohar}},
  \bibnamefont{and} \bibinfo{author}{\bibfnamefont{W.~J.}
  \bibnamefont{Waalewijn}}, \bibinfo{journal}{JHEP}
  \textbf{\bibinfo{volume}{05}}, \bibinfo{pages}{106} (\bibinfo{year}{2018}),
  \eprint{1803.06347}.

\end{thebibliography}

\end{document}